\documentclass[9pt]{article}
\usepackage{graphicx}
\usepackage{amsmath}

\begin{document}


\centerline{\bf Glueball filled with the quark field as a model of nucleon}

\bigskip

\centerline{Vladimir Dzhunushaliev}

\medskip

\centerline{\small by Senior Associate}

\centerline{\small Dept. Phys. and Microel. Engineer.,
}

\centerline{\small KRSU, Bishkek, Kievskaya Str. 44,}
\centerline{\small 720021, Kyrgyz Republic}

\centerline{\small the Abdus Salam ICTP}

\centerline{\small\tt dzhun@hotmail.kg}

\bigskip

\centerline{\small \bf Received September 9, 2004}

\medskip

\begin{abstract}
On the basis of the non-perturbative Heisenberg's quantization
scheme and using some simplifications and assumptions the
reduction from the gluon-quark Lagrangian to a scalar-fermion
Lagrangian is made. The corresponding field equations have a
regular spherically symmetric solution which can be interpreted as
a simplified model of nucleon. Some properties of such a
``nucleon'' are discussed: the presence of a mass gap and spin
$\neq \hbar /2$. The mass gap describes the difference between the
``nucleon'' and the ``nucleon'' without quark field (a scalar
glueball). The presence of spin $\neq \hbar /2$ can be connected 
with the spin problem of nucleon.
\end{abstract}


\section{Introduction}

One can assume that removing the quarks from a nucleon gives us
a glueball. In this case one of the methods to prove a glueball
model is by filling up it with fermions (quarks) and comparing it with 
nucleon properties. In Ref. \cite{dzhun1} a scalar model of
glueball is presented. This model is based on a non-perturbative
quantization technique which was applied by Heisenberg for a
non-linear spinor field. In our approach we have applied this idea
for quantum chromodynamics. The essence of this non-perturbative
technique is that in the first approximation the field degrees of
freedom ($A_{\mu}^{B}$ gauge potential, $B=1,2,\ldots8$ is the
color indices, $\mu=0,1,2,3$ is the world indices) can be reduced
to scalar fields $\phi^{B}$. The components $\phi^{a}$($a=1,2,3$)
and $\phi^{m}$ ($m=4,5,6,7,8$) of the scalar field have a
different dynamical behavior. The numerical analysis shows that
the corresponding field equations for $\phi^{a,m}$ have regular
solutions only for some values of parameters of the model. The
solution looks like a bag, into which we would like to put quarks
(fermions).

\section{Short description of the model}

In this section we would like to derive the field equations
describing in an approximate manner quantized fermion and SU(3)
gauge fields. Strictly speaking the Heisenberg's quantization
procedure \cite{heis} is based on an infinite set of equations
relating all Green's functions which are similar to
Dayson-Schwinger equations but on the non-perturbative language.
Of course such a set of equations is mathematically extremely hard
for solving and we need for some physical reasonings for a
closure. In the approach presented here we consider only 2 and
4-point Green's functions. In this case we can average the quantum
chromodynamics Lagrangian and derive equations describing 2 and
4-point Green's functions. For this purpose we assume that a 4-point
Green's function can be presented as some sum of the products of
2-point Green's functions, and 2-point Green's function can be
decomposed as the product of scalar fields. The Lagrangian of the
SU(3) gauge field interacting with quarks is
\begin{equation}
  \widehat \mathcal{L} = \widehat \mathcal{L}_g + \widehat
  \mathcal{L}_q =
  - \frac{1}{4} \widehat F^A_{\mu \nu}\widehat F^{A \mu \nu} +
  i \widehat{\bar{\psi}} \gamma^\mu
  \left(
    \partial_\mu + i G \widehat A_\mu
  \right) \widehat \psi -
  M \widehat{\bar{\psi}} \psi
\label{sec1:10}
\end{equation}
where $\widehat \mathcal{L}_{g,q}$ are the gluon and quark
Lagrangians correspondingly; $\widehat F^B_{\mu \nu} = \partial_\mu
\widehat A^B_\nu -
\partial_\nu \widehat A^B_\mu + G f^{BCD} \widehat A^C_\mu \widehat A^D_\nu$
is the field strength operator; $B,C,D = 1, \ldots ,8$ are the
SU(3) color indices; $G$ is the coupling constant; $f^{BCD}$ are
the structure constants for the SU(3) gauge group; $\widehat
A^B_\mu$ is the gauge potential operator; $\widehat A_\mu =
L^B\widehat A^B_\mu$; $L^B$ are the generators of the SU(3) gauge
group; $\widehat \psi$ is the quark field and $M$ is the mass
matrix of quarks. In order to derive equations describing the
quantized field we average this Lagrangian over a quantum state
$\left.\left. \right| Q \right\rangle$
\begin{equation}
    \left\langle Q \left| \widehat \mathcal{L}(x) \right| Q \right\rangle =
    \left\langle \widehat \mathcal{L} \right\rangle =
    \left\langle \widehat \mathcal{L}_g \right\rangle +
    \left\langle \widehat \mathcal{L}_q \right\rangle
\label{sec1:20}
\end{equation}
where
\begin{eqnarray}
  \left\langle \widehat \mathcal{L}_g \right\rangle &=&
  - \left[\frac{1}{2}
    \left\langle
      \left( \partial_\mu \widehat A^B_\nu (x) \right)
      \left( \partial^\mu \widehat A^{B\nu} (x) \right) -
      \left( \partial_\mu \widehat A^B_\nu (x) \right)
      \left( \partial^\nu \widehat A^{B\mu} (x) \right)
    \right\rangle  \right.
\nonumber \\
    &&
    =\frac{1}{2} G f^{BCD}
    \left\langle
      \left( \partial_\mu \widehat A^B_\nu (x)-
      \partial_\nu \widehat A^B_\mu (x)\right)
      \widehat A^{C \mu} (x)\widehat A^{D \nu}(x)
    \right\rangle
\nonumber \\
    &&
    +\left.\frac{1}{4}G^2 f^{BC_1D_1} f^{BC_2D_2}
    \left\langle
      \widehat A^{C_1}_\mu (x)\widehat A^{D_1}_\nu (x)
      \widehat A^{C_2 \mu} (x)\widehat A^{D_2\nu} (x)
    \right\rangle \right] ,
\label{sec1:30}\\
  \left\langle \widehat \mathcal{L}_q \right\rangle &=&
  i \widehat{\bar{\psi}} \gamma^\mu
  \left(
    \partial_\mu + i G \widehat A_\mu
  \right) \widehat \psi
\label{sec1:40}
\end{eqnarray}
At first we will calculate $\left\langle \widehat \mathcal{L}_g
\right\rangle$ following 
Refs.~\cite{dzhun1,Dzhunushaliev:2003zi}. One can see that
schematically we have the following 2, 3, and 4-point Green's
functions: $\left\langle \left( \partial A \right)^2
\right\rangle$, $\left\langle \left( \partial A \right) A^2
\right\rangle$ and $\left\langle \left( A \right)^4\right\rangle$.
We suppose that the odd Green's functions can be written as the
following product:
\begin{equation}
    \left\langle
      \widehat A^B_\alpha (x)\widehat A^C_\beta (y)\widehat A^D_\gamma (z)
    \right\rangle \approx
    \left\langle
      \widehat A^B_\alpha (x)\widehat A^C_\beta (y)
    \right\rangle
    \left\langle
      \widehat A^D_\gamma (z)
    \right\rangle + \text{(other permutations)}
    = 0
\label{sec1:50}
\end{equation}
as we suppose that $\langle \widehat A^B_\alpha (x) \rangle = 0$.
Further we suppose that a 2-point Green's function can be presented
in the so-called one-function approximation
\cite{Dzhunushaliev:2003zi} as
\begin{equation}
    \left\langle
      \widehat A^B_\alpha (x) \widehat A^C_\beta (y)
    \right\rangle =
    \mathcal{G}^{BC}_{\alpha \beta} (x,y)
  \approx
    -\eta_{\alpha \beta} f^{BAD} f^{CAE} \phi^D (x) \phi^E(y)
\label{sec1:60}
\end{equation}
where $\phi^A(x)$ is the scalar field which describes the 2-point
Green's function. These two assumptions are similar to the quantum
harmonic oscillator where $\left\langle x \right\rangle = 0$ but
$\left\langle x^2 \right\rangle \neq 0$. The 4-point Green's
function can be written in a one-function approximation as in the
symmetrized product of corresponding two 2-point Green's functions
\begin{equation}
\begin{split}
    &\left\langle
      \widehat A^B_\alpha (x) \widehat A^C_\beta (y)
      \widehat A^D_\gamma (z) \widehat A^R_\delta (u)
    \right\rangle \approx
    \left\langle
      \widehat A^B_\alpha (x) \widehat A^C_\beta (y)
    \right\rangle
    \left\langle
      \widehat A^D_\gamma (z) \widehat A^R_\delta (u)
    \right\rangle  \\
    &+\left\langle
      \widehat A^B_\alpha (x) \widehat A^D_\gamma (z)
    \right\rangle
    \left\langle
      \widehat A^C_\beta (y) \widehat A^R_\delta (u)
    \right\rangle +
    \left\langle
      \widehat A^B_\alpha (x) \widehat A^R_\delta (u)
    \right\rangle
    \left\langle
      \widehat A^C_\beta (y) \widehat A^D_\gamma (z)
    \right\rangle .
\end{split}
\label{sec1:70}
\end{equation}
Taking into account these expressions for the 2, 3, and 4-point
Green's functions we can derive the effective Lagrangian $\mathcal
{L}_{\phi} = \left\langle \widehat \mathcal {L}_g \right\rangle$
for the scalar field $\phi^A$ which describes 2 and 4-point
Green's functions
\begin{equation}
\begin{split}
    &\mathcal {L}_{\phi} =
    \left\langle \mathcal {L}_{g} \right\rangle \approx
        \frac{4}{G^2}
        \left[
        \frac{1}{2}\left( \partial_\mu \phi^A \right)^2 -
        \frac{\lambda_2}{4}
        \left[ \phi^a \phi^a - \phi^a_0 \phi^a_0
        \right]^2 +
        \frac{\lambda_2}{4} \left( \phi^a_0 \phi^a_0 \right)^2
    \right.\\
    &
    -\left.
    \frac{\lambda_1}{4}
    \left[ \phi^m \phi^m - \phi^m_0 \phi^m_0
    \right]^2 +
    \frac{\lambda_2}{4} \left( \phi^m_0 \phi^m_0 \right)^2
     - \left( \phi^a \phi^a \right) \left( \phi^m \phi^m \right)
    \right]
\end{split}
\label{sec1:80}
\end{equation}
where the indices $a=1,2,3$ are SU(2) indices and $m=4,5,6,7,8$ are
the coset SU(3)/SU(2) indices; $\phi^A_0$ are some constants. We
can add the term $-\lambda_2(\phi^m_0 \phi^m_0)^2/4$, which is 
inessential for the dynamics (but essential for the finiteness of
the energy), and the Lagrangian becomes
\begin{equation}
\begin{split}
    &\mathcal {L}_{\phi} =
    \left\langle \mathcal {L}_{g} \right\rangle =
        \frac{4}{G^2}
        \left[
        \frac{1}{2}\left( \partial_\mu \phi^A \right)^2 -
        \frac{\lambda_2}{4}
        \left[ \phi^a \phi^a - \phi^a_0 \phi^a_0
        \right]^2
    \right.\\
    &
    -\left.
        \frac{\lambda_1}{4} \phi^m \phi^m
        \left[ \phi^m \phi^m - 2 \phi^m_0 \phi^m_0
        \right]^2 -
        \left( \phi^a \phi^a \right) \left( \phi^m \phi^m \right)
    \right] .
\end{split}
\label{sec1:85}
\end{equation}
The next step is calculation of $\left\langle \mathcal {L}_{q}
\right\rangle$,
\begin{equation}
   \left\langle \widehat \mathcal{L}_q \right\rangle =
   \left\langle \widehat{\bar{\psi}}
   \biggl(
     i \gamma^\mu \partial_\mu - M
   \biggl)
   \widehat{\psi}
   \right\rangle -
   G \left\langle
     \widehat{\bar{\psi}} \gamma^\mu \widehat A_\mu \widehat{\psi}
   \right\rangle .
\label{sec1:90}
\end{equation}
We see that the fermion (quark) field $\psi$ does not have any
strong self-interaction. Therefore we can suppose that in the
first approximation the dynamical behavior of the quantized quark
field $\psi$ is similar to the dynamics of classical field
\begin{equation}
   \left\langle \widehat \mathcal{L}_q \right\rangle \approx
   \bar{\psi}
   \biggl(
     i \gamma^\mu \partial_\mu - M
   \biggl)
   \psi -
   G \bar{\psi} \gamma^\mu \left\langle
     \widehat A_\mu
   \right\rangle \psi .
\label{sec1:100}
\end{equation}
Here, we have to note that according to our previous assumptions
$\left\langle A^B_\mu \right\rangle = 0$ and consequently the
second term in Eq.~\eqref{sec1:100} is zero but the interaction
between gluon and quark fields \textit{must} exist. To
introduce the interaction between $\phi^A$ and $\psi$ we have to
do some physical assumptions about this interaction and insert
this term into Lagrangian by hand. For example, one can assume
that the quark field \emph{interacts with fluctuations of the
gluon field},
\begin{equation}
   \mathcal{L}_{int} = - \alpha '
   \left(
    \bar{\psi} \psi
   \right)
   \left(
    \widehat A^B_\mu \widehat A^{B\mu}
   \right) =
   \alpha
   \left(
    \bar{\psi} \psi
   \right)
   \left(
    \phi^A \phi^A
   \right)
\label{sec1:110}
\end{equation}
where for the redefinition of $\alpha '$ we have used Eq.
\eqref{sec1:60}. This assumption is equivalent to cutting off the
infinite set of equations connecting all Green's functions for
exactly quantized gluon and quark fields. Thus the quark
Lagrangian is
\begin{equation}
   \left\langle \widehat \mathcal{L}_q \right\rangle \approx
   \bar{\psi}
   \biggl[
     i \gamma^\mu \partial_\mu -
     \Bigl(
        M - \alpha \phi^A \phi^A
     \Bigl)
   \biggl] \psi.
\label{sec1:120}
\end{equation}
Hereafter, we omit flavor indices to simplify notation. Finally we
have the following effective Lagrangian describing the
non-perturbative quantized qluon and quark fields interacting
amongst themselves
\begin{equation}
\begin{split}
    \left\langle \widehat \mathcal{L} \right\rangle =
    \frac{4}{G^2}
    \left[
        \frac{1}{2}\left( \partial_\mu \phi^A \right)^2 -
        \frac{\lambda_2}{4}
        \left[ \phi^a \phi^a - \phi^a_0 \phi^a_0
        \right] \right.\\ \left.
        -\frac{\lambda_1}{4} \phi^m \phi^m
        \left[ \phi^m \phi^m - 2 \phi^m_0 \phi^m_0
        \right] -
        \left( \phi^a \phi^a \right) \left( \phi^m \phi^m \right)
    \right] \\
   +\bar{\psi}
   \biggl[
     i \gamma^\mu \partial_\mu -
     \Bigl(
        M - \alpha \phi^A \phi^A
     \Bigl)
   \biggl] \psi
\label{sec1:125}
\end{split}
\end{equation}
with the following field equations
\begin{eqnarray}
  \nabla^\mu \nabla_\mu \phi^a = - \phi^a
  \left[
    2 \phi^m \phi^m + \lambda_2
    \left(
        \phi^a \phi^a - \phi^a_0 \phi^a_0
    \right) -
    \frac{\alpha G^2}{2} \bar{\psi} \psi
  \right],
\label{sec1:130}\\
  \nabla^\mu \nabla_\mu \phi^m = - \phi^m
  \left[
    2 \phi^a \phi^a + \lambda_1
    \left(
        \phi^m \phi^m - \phi^m_0 \phi^m_0
    \right) -
    \frac{\alpha G^2}{2} \bar{\psi} \psi
  \right],
\label{sec1:140}\\
    \left[
        i \gamma^\mu \partial_\mu -
        \left(
            M - \alpha \phi^A \phi^A
        \right)
    \right] \psi = 0.
\label{sec1:150}
\end{eqnarray}

\section{Initial equations}

In this section we would like to present the solution which later
we will view as a simplified model of the nucleon.  We search
solution in the following spherically symmetric form:
\begin{eqnarray}
    \phi^a (r) &=& \frac{\phi(r)}{\sqrt{6}} , \quad a=1,2,3 ;
\label{sec2:10}\\
  \phi^m (r) &=& \frac{f(r)}{\sqrt{10}} , \quad m=4,5,6,7,8 ;
\label{sec2:20}\\
  \psi \left( t,r, \theta, \varphi \right) &=& \mathrm e^{-iEt}
  \left(
    \begin{array}{l}
        h(r) \\
        0 \\
        g(r) \cos \theta \\
        g(r) \sin \theta \mathrm e^{-i\varphi} \\
    \end{array}
  \right)
\label{sec2:30}
\end{eqnarray}
where $r, \theta , \varphi$ are spherical coordinates. Once again
we would like to remind that in our approach we suppose that the
components of scalar fields with different indices $\phi^a,
a=1,2,3$ and $\phi^m, m=4,5,6,7$ have different dynamical
behavior. After s
ubstituting Eqs.~\eqref{sec2:10}-\eqref{sec2:30} into
Eqs.~\eqref{sec1:130}-\eqref{sec1:150} gives
\begin{eqnarray}
    \phi'' + \frac{2}{r} \phi' &=& \phi
    \left[
      f^2 + \lambda_2 \left( \phi^2 - m^2 \right) -
      \frac{\alpha G^2}{2}
      \left(
        h^2 - g^2
      \right)
    \right],
\label{sec2:40}\\
    f'' + \frac{2}{r} f' &=& f
    \left[
      \phi^2 + \lambda_1 \left( f^2 - \mu^2 \right)-
      \frac{\alpha G^2}{2}
      \left(
        h^2 - g^2
      \right)
    \right],
\label{sec2:50}\\
    h' + \frac{2}{r} h &=& g
    \left[
        E + M - \frac{\alpha}{2}
        \left(
            f^2 + \phi^2
        \right)
    \right],
\label{sec2:60}\\
    g' &=& - h
    \left[
        E - M + \frac{\alpha}{2}
        \left(
            f^2 + \phi^2
        \right)
    \right]
\label{sec2:70}
\end{eqnarray}
where $2\phi_0^a \phi_0^a = m^2$ and $2 \phi_0^m \phi_0^m =
\mu^2$; $m, \mu$ are some constants which will be calculated by
solving Eqs.~\eqref{sec2:40}-\eqref{sec2:70}; and constants
$\lambda_{1,2}$ are redefined $\lambda_{1,2} / 2 \rightarrow
\lambda_{1,2}$. We redefine $\phi(r)/\phi(0) \rightarrow \phi(x)$,
$f(r)/\phi(0) \rightarrow f(x)$, $m/\phi(0) \rightarrow m$,
$\mu/\phi(0) \rightarrow \mu$, $h(r)/\phi^{3/2}(0) \rightarrow
h(x)$, $g(r)/\phi^{3/2}(0) \rightarrow g(x)$, $\alpha /\phi(0)
\rightarrow \alpha$, $E/\phi(0) \rightarrow E$, $M/\phi(0)
\rightarrow M$ and introduce the dimensionless radius
$x=r\phi(0)$. After this we have the following set of equations:
\begin{eqnarray}
    \phi'' + \frac{2}{x} \phi' &=& \phi
    \left[
      f^2 + \lambda_2 \left( \phi^2 - m^2 \right) -
      \frac{\alpha G^2}{2}
      \left(
        h^2 - g^2
      \right)
    \right],
\label{sec2:80}\\
    f'' + \frac{2}{x} f' &=& f
    \left[
      \phi^2 + \lambda_1 \left( f^2 - \mu^2 \right)-
      \frac{\alpha G^2}{2}
      \left(
        h^2 - g^2
      \right)
    \right],
\label{sec2:90}\\
    h' + \frac{2}{x} h &=& g
    \left[
        E + M - \frac{\alpha}{2}
        \left(
            f^2 + \phi^2
        \right)
    \right],
\label{sec2:100}\\
    g' &=& - h
    \left[
        E - M + \frac{\alpha}{2}
        \left(
            f^2 + \phi^2
        \right)
    \right]
\label{sec2:110}
\end{eqnarray}
where $d(\cdots)/dx = (\cdots)'$. Evidently these equations are too
complicated to be solved analytically. Preliminary numerical
investigations show us that the set
\eqref{sec2:80}-\eqref{sec2:110} does not have regular solutions
for arbitrary choice of the parameters $m$, $\mu$, and $E$. We
will solve equations \eqref{sec2:80}-\eqref{sec2:110} as a
nonlinear eigenvalue problem for eigenstates $\phi(x), f(x), h(x),
g(x)$ and eigenvalues $m, \mu, E$, i.e. we have to find parameters
$m, \mu, E$ to provide existence of regular functions $\phi(r)$,
$f(r)$, $h(r)$ and $g(r)$. We will search for the regular solution
under the following boundary conditions
\begin{eqnarray}
    \phi(0) &=& 1, \quad \phi(\infty) = m ;
\label{sec2:120}\\
  f(0) &=& f_0, \quad f(\infty) = 0 ;
\label{sec2:130}\\
    h(0) &=& 0;
\label{sec2:140}\\
    g(0) &=& g_0.
\label{sec2:150}
\end{eqnarray}
The boundary conditions for $\phi(x)$ and $f(x)$ allow us to say
that these functions resemble interacting kink and soliton.
\par
For the regular solution which will be presented below we assign
the following values:
 $\phi(0) = 1.0; f(0) = \sqrt{0.6}; g(0) =
0.1; G = 2.0; M = 0.2; \lambda_1 = 1.0; \lambda_2 = 0.1; \alpha =
1.0$.

\section{Numerical solution}

In accordance Ref.~\cite{dzhun1}, we use the following
numerical method for solving
Eqs.~\eqref{sec2:80}-\eqref{sec2:110}: we take a zero
approximation for the functions $f(x), h(x), g(x)$ (which are
$f_0(x) = 0.6/\cosh^2(x/4)$ and $h_0(x)=g_0(x)=0$) and solve
equation \eqref{sec2:80} in the following form:
\begin{equation}
    \phi_1'' + \frac{2}{x} \phi_1 ' = \phi_1
    \left[
      f^2_0 + \lambda_2 \left( \phi^2_1 - m^2_1 \right)
    \right]
\label{sec3:10}
\end{equation}
where $m_1$ is the first approximation for the parameter $m$, the
boundary conditions are \eqref{sec2:120} and the function
$\phi_1(x)$ is the first approximation for the function $\phi(x)$;
$\phi_1(x)$ and $m_1$  are the eigenfunction and the eigenvalue
for Eq.~\eqref{sec2:80} correspondingly, with the boundary
conditions \eqref{sec2:120}. Obtaining the regular solution
$\phi_1(x)$ we can substitute it into Eq.~\eqref{sec2:90} with
zero approximations $h_0(x)=g_0(x)=0$ and solve the equation
\begin{equation}
    f_1'' + \frac{2}{x} f_1' = f_1
    \left[
      \phi^2_1 + \lambda_1 \left( f^2_1 - \mu^2_1 \right)
    \right]
\label{sec3:20}
\end{equation}
with the boundary conditions \eqref{sec2:130}. $f_1(x)$ and
$\mu_1$  are correspondingly the eigenfunction and the eigenvalue
for Eq.\eqref{sec2:90}, with the boundary conditions
\eqref{sec2:130}.
\par
Further having the first approximations $\phi_1(x), f_1(x)$ we can
substitute them into Eqs.~\eqref{sec2:100} and \eqref{sec2:110}
\begin{eqnarray}
    h'_1 + \frac{2}{x} h'_1 &=& g_1
    \left[
        E + M - \frac{\alpha}{2}
        \left(
            f_1^2 + \phi_1^2
        \right)
    \right],
\label{sec3:30}\\
    g_1' &=& - h_1
    \left[
        E - M + \frac{\alpha}{2}
        \left(
            f_1^2 + \phi_1^2
        \right)
    \right]
\label{sec3:40}
\end{eqnarray}
where $h_1(x), g_1(x), E_1$ are eigenfunctions and eigenvalue for
Dirac equations \eqref{sec3:30} and \eqref{sec3:40} with the
potential $\frac{\alpha}{2}(f_1^2 + \phi_1^2)$. Thus having the
first approximations for $\phi_1(x), f_1(x), h_1(x), g_1(x), m_1,
\mu_1, E_1$ one can substitute these quantities into
Eq.~\eqref{sec2:80},
\begin{equation}
    \phi_2'' + \frac{2}{x} \phi_2 ' = \phi_2
    \left[
      f^2_1 + \lambda_2 \left( \phi^2_2 - m^2_1 \right) -
      \frac{\alpha G^2}{2}
      \left(
        h_1^2 - g_1^2
      \right)
    \right].
\label{sec3:50}
\end{equation}
This equation gives us the second approximation $\phi_2(x)$ for
the function $\phi(x)$, and so on. Thus, on the $i^{th}$ step we
have
\begin{eqnarray}
    \phi_i'' + \frac{2}{x} \phi_i ' &=& \phi_i
    \left[
      f^2_{i-1} + \lambda_1 \left( \phi^2_i - m^2_i \right) -
      \frac{\alpha G^2}{2}
      \left(
        h_{i-1}^2 - g_{i-1}^2
      \right)
    \right]
\label{sec3:60}\\
    f_{i}'' + \frac{2}{x} f_{i}' &=& f_{i}
    \left[
      \phi_{i}^2 + \lambda_1 \left( f_{i}^2 - \mu_{i}^2 \right)-
      \frac{\alpha G^2}{2}
      \left(
        h_{i-1}^2 - g_{i-1}^2
      \right)
    \right],
\label{sec3:62}\\
    h_{i}' + \frac{2}{x} h_{i}' &=& g_{i}
    \left[
        E_{i} + M - \frac{\alpha}{2}
        \left(
            f_{i}^2 + \phi_{i}^2
        \right)
    \right],
\label{sec3:64}\\
    g_{i}' &=& - h_{i}
    \left[
        E_{i} - M + \frac{\alpha}{2}
        \left(
            f_{i}^2 + \phi_{i}^2
        \right)
    \right]
\label{sec3:66}
\end{eqnarray}
where the functions $f_{i-1}(x), h_{i-1}(x), g_{i-1}(x)$ are
defined in the preceding step,
\begin{eqnarray}
    f_{i-1}'' + \frac{2}{x} f_{i-1}' &=& f_{i-1}
    \left[
      \phi_{i-1}^2 + \lambda_1 \left( f_{i-1}^2 - \mu_{i-1}^2 \right)-
      \frac{\alpha G^2}{2}
      \left(
        h_{i-2}^2 - g_{i-2}^2
      \right)
    \right],
\label{sec3:70}\\
    h_{i-1}' + \frac{2}{x} h_{i-1}' &=& g_{i-1}
    \left[
        E_{i-1} + M - \frac{\alpha}{2}
        \left(
            f_{i-1}^2 + \phi_{i-1}^2
        \right)
    \right],
\label{sec3:80}\\
    g_{i-1}' &=& - h_{i-1}
    \left[
        E_{i-1} - M + \frac{\alpha}{2}
        \left(
            f_{i-1}^2 + \phi_{i-1}^2
        \right)
    \right]
\label{sec3:90}
\end{eqnarray}
Such procedure has to lead to precise eigenfunctions
\begin{equation}
    \phi^*_{i}(x) \rightarrow \phi^*(x), \quad
    f^*_{i}(x) \rightarrow f^*(x), \quad
    h^*_{i}(x) \rightarrow h^*(x), \quad
    g^*_{i}(x) \rightarrow g^*(x)
\label{sec3:100}
\end{equation}
with precise eigenvalues
\begin{equation}
 m^*_{i} \rightarrow m^*, \quad
 \mu^*_{i} \rightarrow m^*, \quad
 E^*_{i} \rightarrow E^*
\label{sec3:110}
\end{equation}
where $\phi^*_i, f^*_i, h^*_i, g^*_i$ are the regular solutions of
Eqs.~\eqref{sec3:60}-\eqref{sec3:66} which can be regular only for
the values of parameters $m^*_i, \mu^*_i, E^*_i$. For the other
values of parameters $m_i \neq m^*_i, \mu_i \neq \mu^*_i, E_i \neq
E^*_i$ the functions $\phi_i \neq \phi^*_i, f_i \neq f^*_i, h_i
\neq h^*_i, g_i \neq g^*_i$ are singular and imply infinite
energy.

\subsection{More detailed description of the numerical
calculations for every step}

In this section we would like to describe more carefully the
procedure of the numerical solution of each equation of the system
\eqref{sec3:60}-\eqref{sec3:90}.
\par
At first we would like to describe more carefully the numerical
solution of equation \eqref{sec3:10}. For this purpose, we choose
the zero approximation $f_0(x)$ as follows:
\begin{equation}
    f_0(x) = \frac{\sqrt{0.6}}{\cosh^2{\frac{x}{4}}}.
\label{sec3a:10}
\end{equation}
Typical solutions for the arbitrary value of $m_1$ are presented
in Fig.~\ref{fig:phi-sing}. We see that when $m_1 < m_1^*$
({$m_1^*$ is an eigenvalue which gives us an eigenfunction
$\phi^*_1(x)$) the solution $\phi_1(x)$ is singular and closer to the
singularity the equation has the form
\begin{figure}[h]
  \begin{center}
    \fbox{
    \includegraphics[height=6cm,width=8cm]{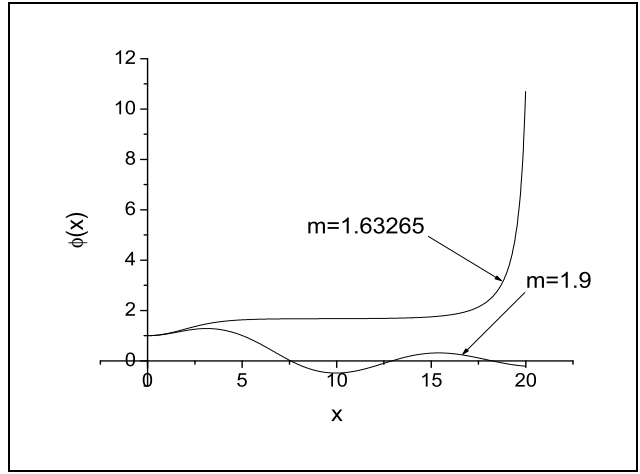}}
    \caption{The typical singular solutions $\phi(x)$
    of Eq.~\eqref{sec3:60}.
    The solution is presented for $i=4$ step.}
    \label{fig:phi-sing}
  \end{center}
\end{figure}
\begin{figure}[h]
  \begin{center}
    \fbox{
    \includegraphics[height=6cm,width=8cm]{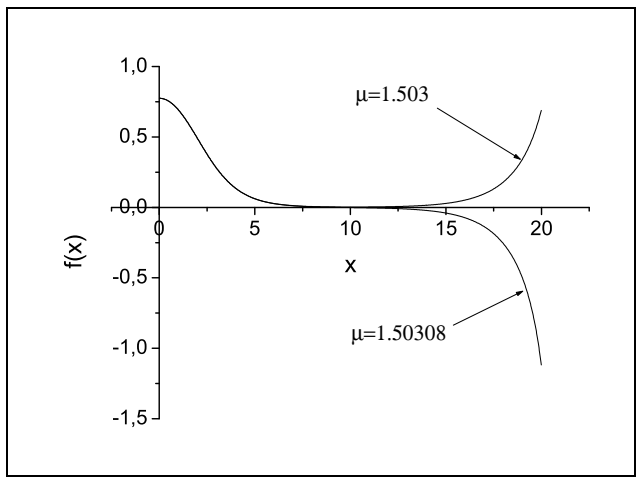}}
    \caption{The typical singular solutions $f(x)$
    of Eq.~\eqref{sec3:62}.
    The solution is presented for $i=4$ step.}
    \label{fig:f-sing}
  \end{center}
\end{figure}
\begin{equation}
    \phi_1'' \approx \lambda_2 \phi_1^3
\label{sec3a:20}
\end{equation}
Consequently the solution is
\begin{equation}
    \phi_1(x) \approx \sqrt{\frac{2}{\lambda_2}} \frac{1}{x_0 - x}
\label{sec3a:30}
\end{equation}
where $x_0$ is some constant depending on $m_1$. On the other
hand, for $m_1 > m_1^*$ the solution is also presented in
Fig.~\ref{fig:phi-sing} and the corresponding asymptotical
equation is
\begin{equation}
    \phi_1''(x) + \frac{2}{x}   \phi_1' \approx -
    \left( \lambda_2 m^2 \right) \phi_1
\label{sec3a:40}
\end{equation}
which has the following solution:
\begin{equation}
    \phi_1(x) \approx \phi_\infty
    \frac{\sin{\left(x \sqrt{\lambda_2 m^2} + \alpha\right)}}{x}
\label{sec3a:50}
\end{equation}
where $\phi_\infty$ and $\alpha$ are some constants. This allows
us to assert that there is an eigenvalue $m^*_1$ for which an
eigenfunction exists. This is presented in
Fig.~\ref{fig:phi-reg}, to some degree of accuracy.
\par
For the value $m^*_1$ equation \eqref{sec3:10} has the
following asymptotical behavior:
\begin{equation}
    \phi_1''(x) + \frac{2}{x}   \phi_1' \approx
    2 \lambda_2 \left( m^*_1 \right)^2
    \left( \phi_1 - m^*_1 \right)
\label{sec3a:60}
\end{equation}
and the corresponding asymptotical solution is
\begin{equation}
    \phi_1(x) \approx m^*_1 + \beta_1
    \frac{e^{-x \sqrt{2
    \lambda_2 \left( m^*_1 \right)^2}}}{x}
\label{sec3a:70}
\end{equation}
where $\beta_1$ is some constant.
\par
The next step is finding the first approximation for the $f_1(x)$
function. The corresponding equation is
\begin{equation}
  f_1'' + \frac{2}{x} f' = f_1
  \left[
    \phi_1^2 + \lambda_1 \left( f_1^2 - \mu^2_1 \right)
  \right].
\label{sec3a:80}
\end{equation}
The numerical investigation shows that for arbitrary $\mu$
there are two different singular solutions which are presented in
Fig.~\ref{fig:f-sing}. Analogously, the singular behavior of the
solution $f_1(x)$ is
\begin{eqnarray}
    f_1(x) &\approx & \sqrt{\frac{2}{\lambda_1}} \frac{1}{x - x_0}
    \quad \text {by} \quad \mu_1 < \mu^*_1 ,
\label{sec3a:90}\\
  f_1(x) &\approx & - \sqrt{\frac{2}{\lambda_1}} \frac{1}{x - x_0}
    \quad \text {by} \quad \mu_1 > \mu^*_1 .
\label{sec3a:100}
\end{eqnarray}
Evidently, we can suppose that there is the eigenfunction
$f^*_1(x)$ for the eigenvalue $\mu_1 = \mu^*_1$ with the following
asymptotical behavior:
\begin{equation}
    f^*_1(x) \approx f_{\infty ,1}
    \frac{e^{- x \sqrt{\left( m_0^* \right)^2 -
    \lambda_2 \left( \mu_1^* \right)^2}}}{x}
\label{sec3a:110}
\end{equation}
where $f_{\infty ,1}$ is some parameter.
\par
The next step is the substitution of the first approximations
$\phi^*_1(x), f^*_1(x)$ into Eqs.~\eqref{sec3:64} and
\eqref{sec3:66}, to find eigenfunctions $h^*_1(x), g^*_1(x)$
corresponding to the eigenvalue $E^*_1$,
\begin{eqnarray}
  h_1' + \frac{2}{x}h &=& g_1
  \left[
    E_1 + M - \frac{\alpha}{2}
    \left[
        \left( f_1^* \right)^2 + \left( \phi_1^* \right)^2
    \right]
  \right] ,
\label{sec3a:120}\\
  g_1 &=& h_1
  \left[
    - E_1 + M - \frac{\alpha}{2}
    \left[
        \left( f_1^* \right)^2 + \left( \phi_1^* \right)^2
    \right]
  \right] .
\label{sec3a:130}
\end{eqnarray}
The numerical solutions of these equations are the same as that of
the Dirac equation for a quantum particle with spin in a potential
hole. In our case the hole is formed by the gluon fields: $V_{eff}
= \alpha (\phi^2 + f^2)/2$. The typical singular solutions of these
equations are presented in Figs.~\ref{fig:h-sing} and
\ref{fig:g-sing}.
\begin{figure}[h]
  \begin{center}
    \fbox{
    \includegraphics[height=6cm,width=8cm]{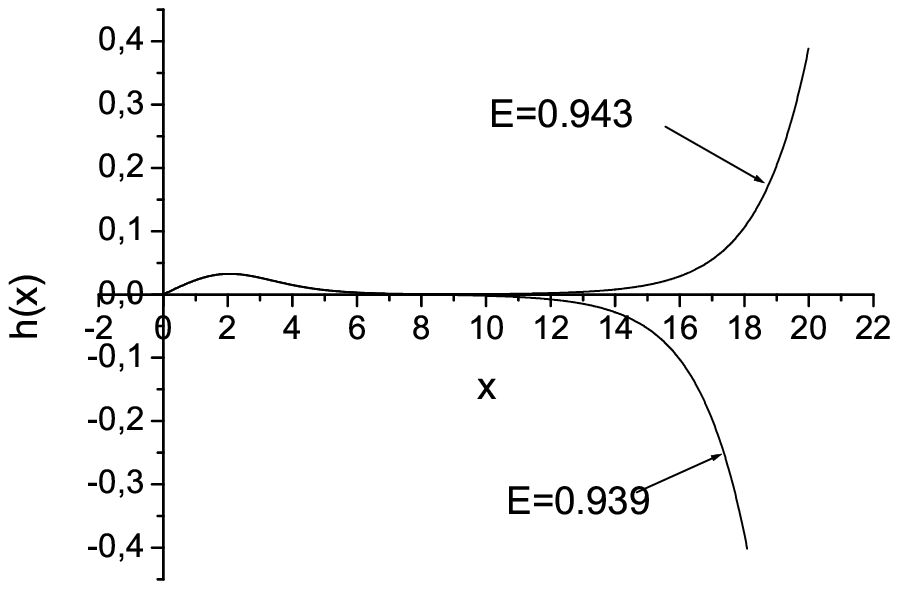}}
    \caption{The typical singular solutions $h(x)$ of
    Eqs.~\eqref{sec3:64} and \eqref{sec3:66}.}
    \label{fig:h-sing}
  \end{center}
\end{figure}
\begin{figure}[h]
  \begin{center}
    \fbox{
    \includegraphics[height=6cm,width=8cm]{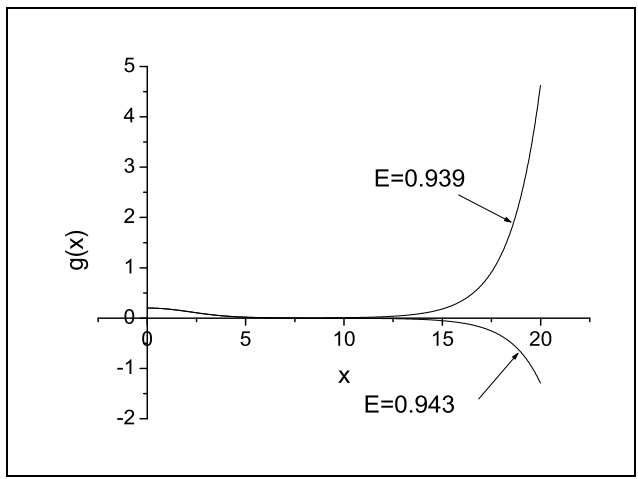}}
    \caption{The typical singular solutions $g(x)$ of
    Eqs.~\eqref{sec3:64} and \eqref{sec3:66}.}
    \label{fig:g-sing}
  \end{center}
\end{figure}
\par
Further, the eigenfunctions $f^*_1(x), h^*_1(x), g^*_1(x)$ will be
substituted into equation \eqref{sec3:60} to find the
eigenfunction $\phi^*_2(x)$ which corresponds to the eigenvalue
$m_2 = m^*_2$, and so on.
\par
The result of these calculations is presented in
Figs.~\ref{fig:phi-reg}, \ref{fig:f-reg}, \ref{fig:h-reg},
\ref{fig:g-reg} and Table \ref{table1}. One can see that there is
the convergence: $\phi^*_i(x) \rightarrow \phi^*(x)$, $f^*_i(x)
\rightarrow f^*(x)$, $h^*_i(x) \rightarrow h^*(x)$, $g^*_i(x)
\rightarrow g^*(x)$, $m^*_i \rightarrow m^*$, $\mu^*_i \rightarrow
\mu^*$ and $E^*_i \rightarrow E^*$, where $f^*(x), \phi^*(x),
h^*(x), g^*(x),$ are the eigenfunctions and $m^*, \mu^*, E^*$ are
the eigenvalues of the nonlinear eigenvalue problem
\eqref{sec2:80}-\eqref{sec2:110}.
\par
Finally we would like to mention that the presented solution
allows us to avoid the Derrick's Theorem \cite{derrick} which
forbids the existence of regular solutions for scalar fields. One
of the necessary conditions of this Theorem is that the solution
at infinity tends to a global minimum of the potential
energy but in our case the solution tends to a local minimum
$\phi^* \rightarrow m^*, f^* \rightarrow 0$ (the global minimum is
$\phi = m, f = \mu$).
\begin{table}[h]
    \begin{center}
        \begin{tabular}{|c|l|l|l|l|}\hline
          i & 1 & 2& 3 & 4 \\ \hline
            $m^*_i$ & 1.837435109502 & 1.61449115 & 1.6296365
            & 1.63265546 \\ \hline
            $\mu^*_i$ & 1.4921052897028195 & 1.504847 & 1.5022397
            & 1.5030292 \\ \hline
            $E^*_i$ & 0.932603482 & 0.942702 & 0.939357
            & 0.940497 \\ \hline
        \end{tabular}
    \end{center}
    \caption{The iterative parameters $m^*_i$, $\mu^*_i$ and $E^*_i$.}
    \label{table1}
\end{table}

\begin{figure}[h]
  \begin{center}
    \fbox{
    \includegraphics[height=6cm,width=8cm]{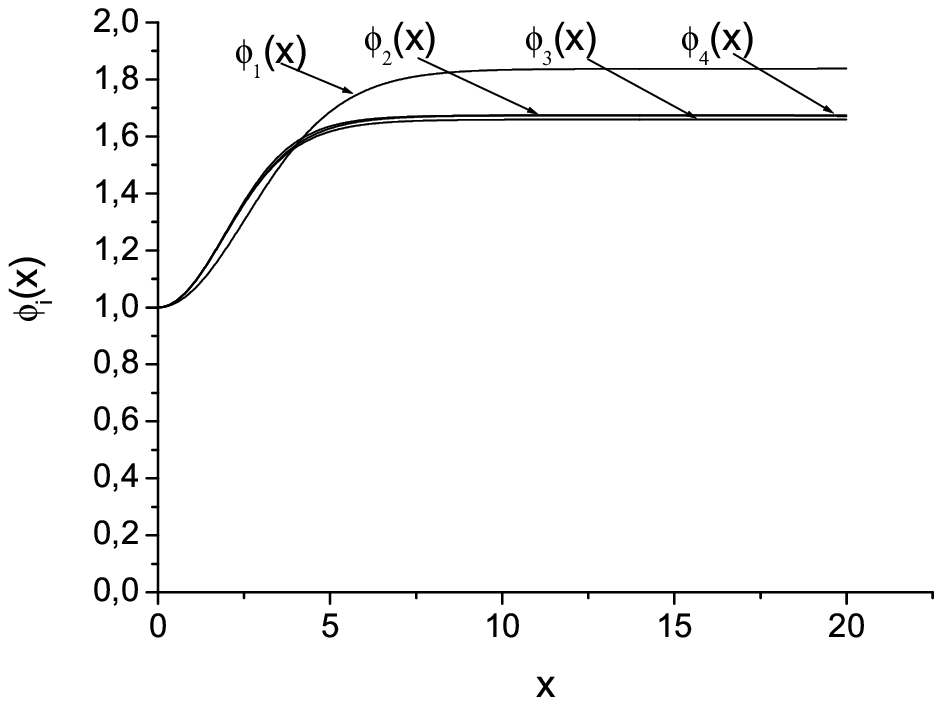}}
    \caption{The iterative functions $\phi_{1,2,3,4}(x)$.}
    \label{fig:phi-reg}
  \end{center}
\end{figure}

\begin{figure}[h]
  \begin{center}
    \fbox{
    \includegraphics[height=6cm,width=8cm]{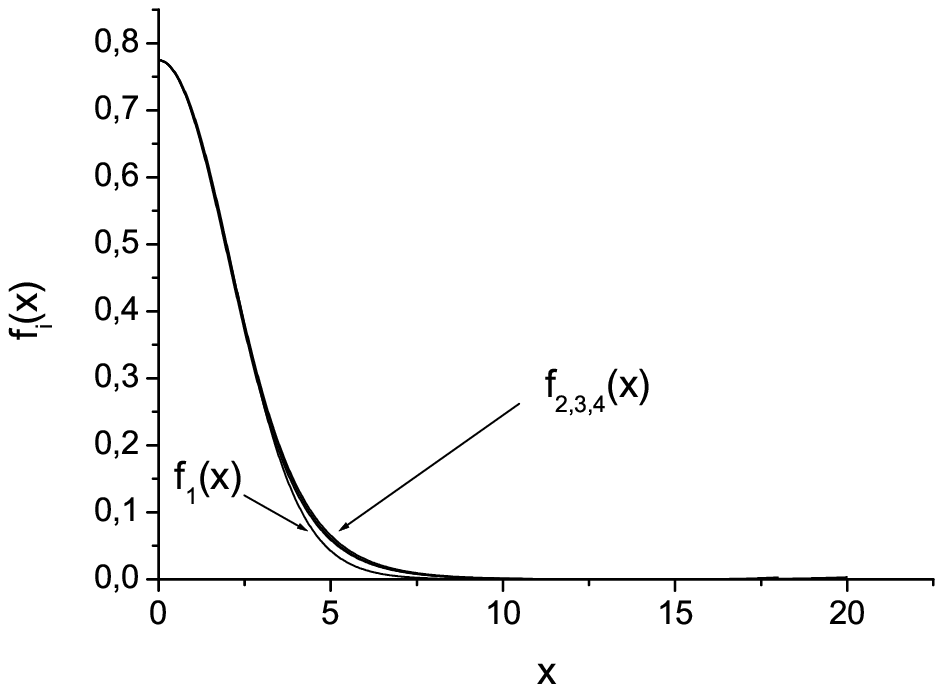}}
    \caption{The iterative functions $f_{1,2,3,4}(x)$.}
    \label{fig:f-reg}
  \end{center}
\end{figure}

\begin{figure}[h]
  \begin{center}
    \fbox{
    \includegraphics[height=6cm,width=8cm]{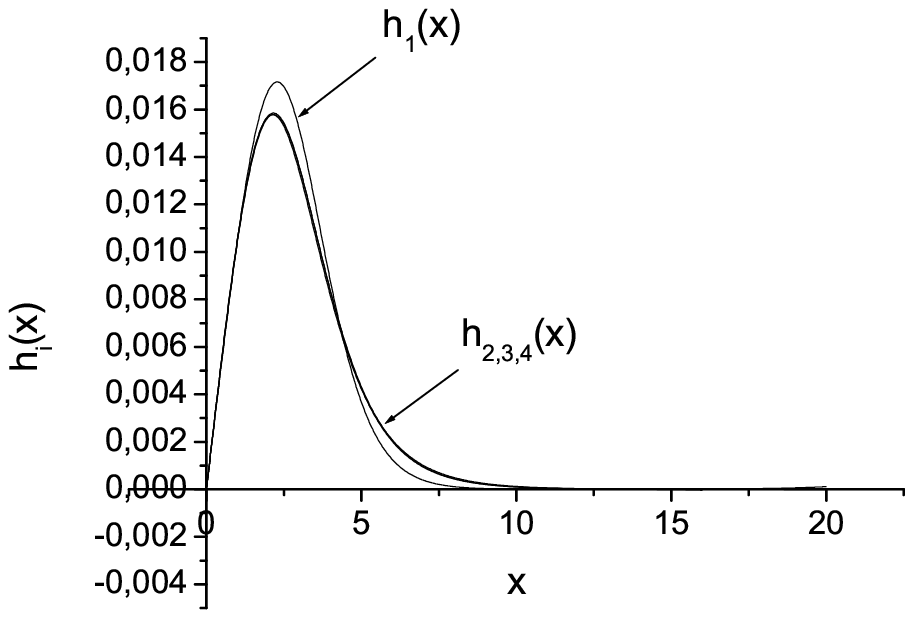}}
    \caption{The iterative functions $h_{1,2,3,4}(x)$.}
    \label{fig:h-reg}
  \end{center}
\end{figure}
\begin{figure}[h]
  \begin{center}
    \fbox{
    \includegraphics[height=6cm,width=8cm]{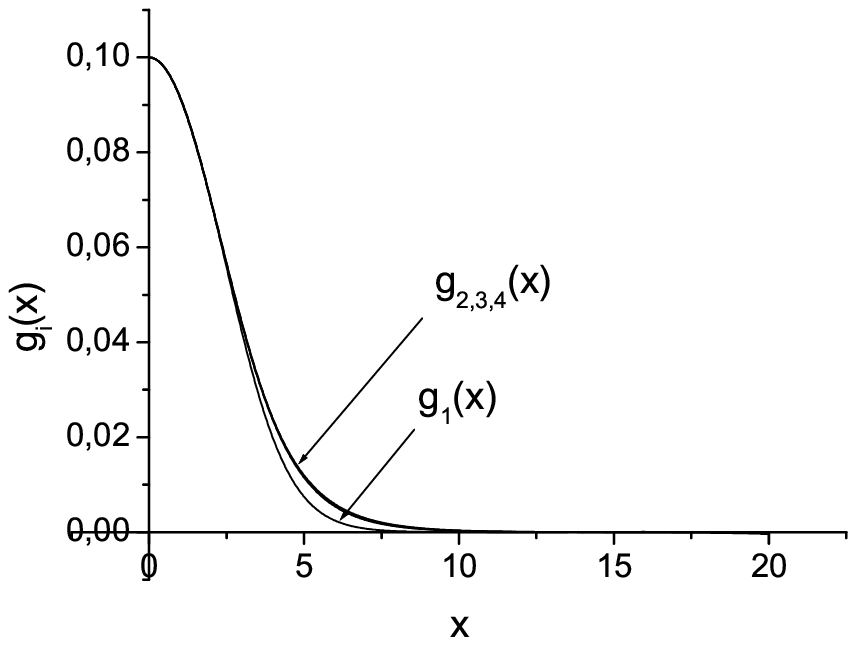}}
    \caption{The iterative functions $g_{1,2,3,4}(x)$.}
    \label{fig:g-reg}
  \end{center}
\end{figure}

\section{Properties of the solution}

In this section we would like to describe properties of the
derived solution. It is easy to see that the asymptotical behavior
of the regular solution is
\begin{eqnarray}
    \phi^*(x) & \approx & m^* + \phi_\infty   \frac{\exp\{-x \sqrt{2
    \lambda_2 \left( m^* \right)^2}\}}{x}
\label{sec5:10}\\
    f^*(x) & \approx & f_\infty
    \frac{\exp\{- x \sqrt{\left( m^* \right)^2 - \lambda_1 \left( \mu^* \right)^2}\}}{x}
\label{sec5:20}\\
    h^*(x) & \approx & h_\infty
    \frac{\exp\{-x \sqrt{\left( M -
    \frac{\alpha \left( m^* \right)^2}{2} \right)^2 - \left( E^* \right)^2}\}}{x}
\label{sec5:22}\\
    g^*(x) & \approx & g_\infty
    \frac{\exp\{-x \sqrt{\left( M -
    \frac{\alpha \left( m^* \right)^2}{2} \right)^2 - \left( E^* \right)^2}\}}{x}
\label{sec5:23}
\end{eqnarray}
\begin{equation}
    h_\infty  =  - g_\infty
    \sqrt{\frac{M - {\alpha \left( m^* \right)^2}/{2} + E^*}
    {M - {\alpha \left( m^* \right)^2}/{2} - E^*}}
\label{sec5:24}
\end{equation}
where $m^*, \mu^*, E^*$ are the eigenvalues
derived by the solving Eqs.~\eqref{sec2:80}-\eqref{sec2:110}.
\par
For determination of the corresponding energy we rewrite the
Lagrangian \eqref{sec1:125} in the form
\begin{equation}
    \left\langle \widehat \mathcal{L} \right\rangle =
    \mathcal{L}_{s} + \mathcal{L}_{int} + \mathcal{L}_\psi
\label{sec5:24a}
\end{equation}
where Lagrangians for the scalar field $\mathcal{L}_s$,
interaction $\mathcal{L}_{int}$, and spinor field
$\mathcal{L}_\psi$ are, correspondingly,
\begin{eqnarray}
    \mathcal{L}_{s} &=&
    \frac{4}{G^2}
    \left[
        \frac{1}{2}\left( \partial_\mu \phi^A \right)^2 -
        \frac{\lambda_2}{4}
        \left( \phi^a \phi^a - \phi^a_0 \phi^a_0
        \right)^2
    \right.
\nonumber \\
    &&-\biggl .
        \frac{\lambda_1}{4} \phi^m \phi^m
        \left( \phi^m \phi^m - 2 \phi^m_0 \phi^m_0
        \right) -
        \left( \phi^a\phi^a \right) \left( \phi^m\phi^m \right)
    \biggl ],
\label{sec5:24b}\\
   \mathcal{L}_{int} &=&
   \alpha \left( \phi^A \phi^A \right) \left( \bar{\psi} \psi \right) ,
\label{sec5:24c}\\
   \mathcal{L}_\psi &=&
   \bar{\psi}
   \left (
     i \gamma^\mu \partial_\mu - M
   \right ) \psi .
\label{sec5:24d}
\end{eqnarray}
Then the energy density is found as follows:
\begin{equation}
\begin{split}
    \varepsilon(r) = &
        \frac{4}{G^2}
    \left[
        \frac{1}{2}\left( \partial_t \phi^A \right)^2 +
        \frac{1}{2}\left( \partial_i \phi^A \right)^2 +
        \frac{\lambda_2}{4}
        \left( \phi^a \phi^a - \phi^a_0 \phi^a_0
        \right)^2
    \right.  \\
    &+\biggl .
        \frac{\lambda_1}{4} \phi^m \phi^m
        \left( \phi^m \phi^m - 2 \phi^m_0 \phi^m_0
        \right) +
        \left( \phi^a\phi^a \right) \left( \phi^m\phi^m \right)
    \biggl ] -
    \alpha \left( \phi^A \phi^A \right) \left( \bar{\psi} \psi \right) +
    \varepsilon_{\psi}.
\label{sec5:24e}
\end{split}
\end{equation}
We will determine the energy density of the quark field
$\varepsilon_{\psi}$ from the energy-momentum tensor of $\psi$,
\begin{equation}
    T^{\mu \nu} = \frac{i}{2} g^{\mu \alpha}
    \left(
       \bar{\psi} \gamma^\mu \frac{\partial \psi}{\partial x^\alpha} -
       \frac{\partial \bar \psi}{\partial x^\alpha} \gamma^\mu \psi
    \right).
\label{sec5:24f}
\end{equation}
Using ans\"atz \eqref{sec2:30} we have
\begin{equation}
    \varepsilon_{\psi} = T^{00} =
    E \left( h^2 + g^2 \right).
\label{sec5:24g}
\end{equation}
Therefore
\begin{equation}
\begin{split}
    \varepsilon(r) = \frac{1}{G^2}
    \left\{
        {\left(\phi^*\right)'}^2 + {\left(f^*\right)'}^2 +
        \frac{\lambda_2}{2}
        \biggl[ \left(\phi^*\right)^2 - \left(m^*\right)^2 \biggl] +
        \frac{\lambda_1}{2} \left(f^*\right)^2
        \biggl[ \left(f^*\right)^2 - 2 \left(\mu^*\right)^2 \biggl]
        \right.\\\left.
        + \left(f^*\right)^2 \left(\phi^*\right)^2
    \right\} -
    \frac{\alpha}{2}
    \biggl[ \left(f^*\right)^2 + \left(\phi^*\right)^2 \biggl]
    \biggl[ \left(h^*\right)^2 - \left(g^*\right)^2 \biggl] +
    E^* \biggl[ \left(h^*\right)^2 + \left(g^*\right)^2 \biggl]
\label{sec5:25}
\end{split}
\end{equation}
where $\lambda_{1,2}, m^*$ and $\mu^*$ are redefined according to the
remark after Eq.~\eqref{sec2:70}, and the functions $(\cdots)$ are
the eigenfunctions. Thus the energy is
\begin{equation}
\begin{split}
    W = &4 \pi \phi_0 \int\limits^\infty_0 x^2
    \Biggl\{
      \frac{1}{G^2}
      \left[
        {\left(\phi^*\right)'}^2 + {\left(f^*\right)'}^2 +
        \frac{\lambda_2}{2}
        \biggl[ \left(\phi^*\right)^2 - \left(m^*\right)^2 \biggl]
    \Biggl.
      \right.\\
        &+\left.
        \frac{\lambda_1}{2} \left(f^*\right)^2
        \biggl[ \left(f^*\right)^2 - 2 \left(\mu^*\right)^2 \biggl] +
        \left(f^*\right)^2 \left(\phi^*\right)^2
      \right]
    \\
    &-\Biggl.
      \frac{\alpha }{2}
      \biggl[ \left(f^*\right)^2 + \left(\phi^*\right)^2 \biggl]
      \biggl[ \left(h^*\right)^2 - \left(g^*\right)^2 \biggl] +
      E^* \biggl[ \left(h^*\right)^2 + \left(g^*\right)^2 \biggl]
    \Biggl\} dx \\
    &=4 \pi \phi_0 I_1 \left( \lambda_{1,2}, m^*, \mu^*, E^*, \alpha \right)
\label{sec5:30}
\end{split}
\end{equation}
where the quantities $r, f^*, \phi^*, h^*, g^*,m^*, \mu^*, E^*,
\alpha $ are dimensionless according to the remark after
Eq.~\eqref{sec2:70}. The quantity $\phi(0)^{-1}$ defines the
radius of the object as a consequence of the asymptotical behavior
\eqref{sec5:10}-\eqref{sec5:24}. The profile of the energy density
is presented in Fig.~\ref{fig:energy}. The numerical calculations
for the integral $I_1$ gives
\begin{equation}
\begin{split}
    I_1 = &\int\limits^\infty_0 x^2
    \Biggl\{
      \frac{1}{G^2}
      \left[
        {\left(\phi^*\right)'}^2 + {\left(f^*\right)'}^2 +
        \frac{\lambda_2}{2}
        \biggl[ \left(\phi^*\right)^2 - \left(m^*\right)^2 \biggl]
    \Biggl.
      \right.\\
        &
        =\left.
        \frac{\lambda_1}{2} \left(f^*\right)^2
        \biggl[ \left(f^*\right)^2 - 2 \left(\mu^*\right)^2 \biggl] +
        \left(f^*\right)^2 \left(\phi^*\right)^2
      \right]
    \\
    &-\Biggl.
      \frac{\alpha }{2}
      \biggl[ \left(f^*\right)^2 + \left(\phi^*\right)^2 \biggl]
      \biggl[ \left(h^*\right)^2 - \left(g^*\right)^2 \biggl] +
      E^* \biggl[ \left(h^*\right)^2 + \left(g^*\right)^2 \biggl]
    \Biggl\} dx \approx 0.395 .
\label{sec5:40}
\end{split}
\end{equation}
\begin{figure}[h]
  \begin{center}
    \fbox{
    \includegraphics[height=6cm,width=8cm]{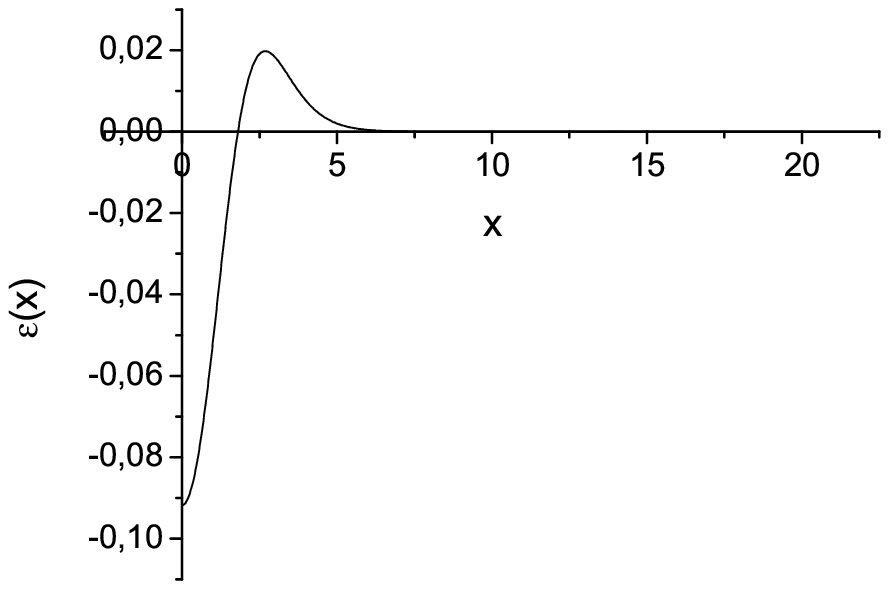}}
    \caption{The profile of the energy density.}
    \label{fig:energy}
  \end{center}
\end{figure}
\par
Another interesting property of the solution is a total angular
momentum density $M_i$. The corresponding operator is
\begin{equation}
    \widehat M_i = \widehat L_i + \widehat s_i
\label{sec5:60}
\end{equation}
where $\widehat L_i$ is the operator of the orbital momentum
density and $\widehat s_i$ is the spin operator density,
\begin{eqnarray}
    \widehat L_i &=& \epsilon_{ijk} x_j
    \left( -i \frac{\partial}{\partial x_k} \right),
\label{sec5:70}\\
    \widehat s_1 &=& i \gamma^2 \gamma^3, \quad
    \widehat s_2 = i \gamma^3 \gamma^1, \quad
    \widehat s_3 = i \gamma^1 \gamma^2
\label{sec5:80}
\end{eqnarray}
where $\gamma^i$ denote Dirac matrices. The calculation of $M_z$ gives
us
\begin{eqnarray}
    \left\langle M_z \right\rangle &=& \frac{1}{2} \psi^* \widehat M_z  \psi =
    \frac{1}{2} \left( h^2 + g^2 \right) ,
\label{sec5:80a}\\
    \left\langle M_x \right\rangle &=&
    \left\langle M_y \right\rangle = 0.
\label{sec5:80b}
\end{eqnarray}
The $z$-projection of the full angular momentum $\mathcal M_z$ is
\begin{equation}
\begin{split}
    \mathcal M_z = &\frac{1}{2} \int \psi^* \widehat M_z  \psi dV =
    \frac{1}{2} \int \left( h^2 + g^2 \right) dV  \\
    &= \frac{1}{2} 4\pi
    \int x^2 \left[ \left( h^* \right)^2 +
    \left( g^* \right)^2 \right] dx =
    \left( \frac{1}{2} \right) 4\pi I_2
\label{sec5:100}
\end{split}
\end{equation}
where $\hbar = 1$ and $I_2 \approx 0.063$.
Most interesting here is that $\psi$ is not a wave function of
a quantum particle but it is the quark field which interacts in a
non-linear way with the scalar fields $\phi^A$ (which after
above-mentioned simplifications present gluon fields).
Consequently, $\psi$ cannot be normalized to unity, i.e.
\begin{equation}
    \int \left( h^2 + g^2 \right) dV \neq 1 .
\label{sec5:110}
\end{equation}
This leads to a very interesting result: the $z$-projection of the
total angular momentum of the solution is $\mathcal M_z < \hbar
/2$\footnote{Note that for $g(0) = 0.2$ we have $\mathcal M_z
> \hbar /2$.}. This can be related to the spin problem of nucleon
which raises experimental and theoretical questions regarding the
contribution of the orbital momentum of the quarks to the total
spin of the nucleon. One can suppose \cite{Singleton:1999eu} that
the full spin of nucleon has two components: one from gluon fields
and another from quark field. Our calculations are in
agreement with this statement: the angular momentum from the quark
field is non-zero and can be $< \hbar /2$. Unfortunately, for the
presented model we cannot calculate the term coming from the
gluon fields since the gluon fields are presented here by the
scalar field which does not have orbital momentum or spin. 
But it can be the aim for 
future investigations. For the simple model presented
here, one can try to tune free parameters of the model,
$(\lambda_{1,2}, M, \alpha , g(0))$ in such a way so as $\mathcal
M_z = \hbar /2$, i.e. $\int \left( h^2 + g^2 \right) dV = 1$, is
fulfilled.

\section{Discussion, conclusion and problems}

In this paper, we have presented a model of nucleon based on the
non-perturbative Heisenberg's quantization technique applied for
the SU(3) gauge Yang-Mills theory plus quark field. We have shown
that under some assumptions and simplifications one can reduce the
gluon-quark Lagrangian to a scalar-fermion Lagrangian. The derived
spherically symmetric solution of corresponding equations can be
considered as a very simplified model of nucleon. Such a
``nucleon'' has a finite mass and spin $<\hbar /2$.
\par
The comparison between the ``nucleon'' and the scalar glueball
(which is ``nucleon'' minus quark field) shows us that there is
\emph{a mass gap} which is relatedyo the presence of the energy
level $E^*$ of the quark field. This value is $E^* \neq 0$ because it
is the \textit{eigenvalue} for the Dirac equation.
\par
Another interesting feature of the ``nucleon'' is that $\mathcal
M_z < \hbar /2$. One can hope that after some complication of the
presented model an additional term from the gluon fields will
appear and the sum of these two terms gives us the spin $=\hbar
/2$. This could resolve the spin problem of the nucleon.
\par
Now we would like to list the assumptions and simplifications which are
necessary for the reduction from the gluon-quark Lagrangian to the
scalar-fermion Lagrangian:
 {\begin{itemize}
    \item
    The infinite set of equations connecting Green's functions are reduced
    using the averaged Lagrangian.
    \item
    This reduction is based on the one-function approximation for the 2- and
    4-point Green's functions.
    \item
    It is assumed that the quark field approximately is classical one.
    \item
    It is assumed that the interaction between gluon and quark fields
    is determined by their mean-square fluctuations:
    $\mathcal L_{int} = \left\langle \left( \widehat{\bar{\psi}} \widehat \psi \right)
    \left( \widehat A^B_\mu \widehat A^B_\mu \right)\right\rangle$.
\end{itemize}
 }
\par
Also, we would like to mention the following problems arising in
the context of the considered model:
\begin{itemize}
    \item
    It is necessary to prove the existence of the presented regular solution.
    Mathematically, this means that one should give an exact ground of the
    presented method of the numerical solution of the non-linear eigenvalue
    problem \eqref{sec2:40}-\eqref{sec2:70}.
    \item
    From the physical point of view, it is necessary to generalize this model
    so that it includes the gluon orbital momentum which can give a contribution
    to spins of the glueball and the nucleon.
    \item
    To compare the presented value of the energy with the mass of nucleon
    it is necessary to tune up free parameters of the model. It is not a simple
    procedure since the preliminary numerical analysis shows us that the regular
    solutions exist not for all values of the parameters.
    \item
    It is interesting to investigate the other possible variants of the
    interaction between gluon and quark fields. In this connection, one can
    mention Ref.~\cite{martens}, in which a model of nucleon is derived on
    the basis of a ``chromo dielectric model''. One can suppose that there
    exists such a gluon-quark interaction that will lead to the
    ``chromo dielectric model''.
\end{itemize}

\section*{Acknowledgment}

I am grateful to the ICTP for financial support and the
invitation to research.


\newpage
\begin{thebibliography}{1}

\bibitem{dzhun1}
V. Dzhunushaliev, ``Scalar model of the glueball'',
hep-ph/0312289, to appear in Found. Phys. Lett.

\bibitem{heis}
W. Heisenberg, \textit{Introduction to the unified field theory of
elementary particles.}, Max - Planck - Institut f\"ur Physik und
Astrophysik, Interscience Publishers London, New York, Sydney,
1966; W. Heisenberg, Nachr. Akad. Wiss. G\"ottingen N8, 111
(1953); W. Heisenberg, Zs. Naturforsch. \textbf{9a}, 292 (1954);
W. Heisenberg, F. Kortel und H. M\"utter, Zs. Naturforsch.
\textbf{10a}, 425 (1955); W. Heisenberg, Zs. f\"ur Phys.
\textbf{144}, 1 (1956); P.~Askali and W. Heisenberg, Zs.
Naturforsch. \textbf{12a}, 177 (1957); W. Heisenberg, Nucl. Phys.
\textbf{4}, 532 (1957); W. Heisenberg, Rev. Mod. Phys.
\textbf{29}, 269 (1957).

\bibitem{Dzhunushaliev:2003zi}
V.~Dzhunushaliev and D.~Singleton,
Mod. Phys. Lett. \textbf{A18}, 2873 (2003).


\bibitem{Singleton:1999eu}
D.~Singleton,
Mod. Phys. Lett. {\bf A16}, 41 (2001).

\bibitem{derrick}
G.H. Derrick, J. Math. Phys. \textbf{5}, 1252 (1964).

\bibitem{martens}
G. Martens, C. Greiner, S. Leupold, and U. Mosel, ``Two- and
three-body color flux tubes in the Chromo Dielectric Model'',
hep-ph/0407215.
\end{thebibliography}
\end{document}